# Title: Convergence of Y chromosome STR haplotypes from different SNP haplogroups compromises accuracy of haplogroup prediction

**Authors**: Chuan-Chao Wang[1], Ling-Xiang Wang[1], Rukesh Shrestha[1], Shaoqing Wen[1], Manfei Zhang[1], Xinzhu Tong[1], Li Jin[1,2,3], Hui Li[1]*

**Affiliations**
1. State Key Laboratory of Genetic Engineering and MOE Key Laboratory of Contemporary Anthropology, School of Life Sciences and Institutes of Biomedical Sciences, Fudan University, Shanghai 200433, China
2. CAS-MPG Partner Institute for Computational Biology, SIBS, CAS, Shanghai, China
3. Institute of Health Sciences, China Medical City, Taizhou, Jiangsu, China
* Correspondence to: lihui.fudan@gmail.com

**Abstract**

Short tandem repeats (STRs) and single nucleotide polymorphisms (SNPs) are two kinds of commonly used markers in Y chromosome studies of forensic and population genetics. There has been increasing interest in the cost saving strategy by using the STR haplotypes to predict SNP haplogroups. However, the convergence of Y chromosome STR haplotypes from different haplogroups might compromise the accuracy of haplogroup prediction. Here, we compared the worldwide Y chromosome lineages at both haplogroup level and haplotype level to search for the possible haplotype similarities among haplogroups. The similar haplotypes between haplogroups B and I2, C1 and E1b1b1, C2 and E1b1a1, H1 and J, L and O3a2c1, O1a and N, O3a1c and O3a2b, and M1 and O3a2 have been found, and those similarities reduce the accuracy of prediction.

**Introduction**

The paternally inherited Y chromosome has been widely used in forensics for personal identification, in anthropology and population genetics to understand origin and migration of human populations, and also in medical and clinical studies. There are two extremely useful markers in Y chromosome, SNP and STR[1,2]. With a very low mutation rate on the order of $3.0 \times 10^{-8}$ mutations/nucleotide/generation[3], SNP markers have been used in constructing a robust phylogeny tree linking all the Y chromosome lineages from world populations[4,5]. Those lineages determined by the pattern of SNPs are called haplogroups. That is to say, we have to genotype an appropriate number of SNPs in order to assign a given Y chromosome to a haplogroup. Compared with SNPs, the mutation rates of STR markers are about four to five orders of magnitude higher. To recognize a Y chromosome, typing STR has advantages of saving time and cost compared with typing SNPs[6]. A set of STR values for an individual is called a haplotype. Because of the disparity in mutation rates between SNP and STR, one SNP haplogroup actually could comprise many STR haplotypes[6]. It is most interesting that STR variability is clustered more by haplogroups than by populations[7,8], which indicates that STR haplotypes could be used to infer the haplogroup information of a given Y chromosome. There has been increasing interest in this cost effective strategy for predicting the haplogroup from a given STR haplotype when SNP data are unavailable. For instance, Vadim Urasin's YPredictor (http://predictor.ydna.ru/), Whit Atheys' haplogroup predictor (http://www.hprg.com/hapest5/) [9,10], and haplogroup classifier of Arizona University[11] have been widely employed in previous studies for haplogroup prediction[12-15].

Although this approach for haplogroup prediction is correct in principle, there are still many ongoing debates about the accuracy of using STR haplotypes in haplogroup assigning [16, 17]. In particular, it has been reported that the number of STRs used in prediction and the available STR-SNP associated reference data have a significant impact on the accuracy of haplogroup prediction[11]. Furthermore, taken the mutation rates of the STRs and the time depth of the haplogroup ramifications into consideration, it is possible to find the same or similar haplotypes from different haplogroups[16]. However, the possible bias caused by the convergence of STR haplotypes in haplogroup prediction has not been discussed before. Here, we use a large amount of worldwide Y chromosome SNP and STR data to address this question.

**Materials and methods**

Altogether, 20403 pieces of Y chromosome data with informative SNP and STR markers have been included in this study (Table S1): unpublished data of 231 East Asian samples from our lab, unpublished data of 101 samples from Genographic Consortium (haplogroup B), and other data retrieved from the literature[4, 17-80]. We renewed the haplogroup names according to the nomenclature of Y Chromosome Consortium and the ISOGG Y-DNA Haplogroup Tree 2013 (http://www.isogg.org/) [4, 5, 81]. Different authors typed different STR markers, which has reduced the feasibility for STR haplotype references. Here, we use the AmpFlSTR® Yfiler™ seventeen Y chromosomal STRs (DYS19, DYS389I, DYS389II, DYS390, DYS391, DYS392, DYS393, DYS385a, DYS385b, DYS438, DYS439, DYS437, DYS448, DYS456, DYS458, DYS635, and YGATAH4) as standard.

Comparisons of STR allele frequency distribution among different haplogroups were performed using the Arlequin ver 3.5 software package (arlecore3513_64bit in Linux)[82]. The Neighbor-joining tree was constructed in MEGA 5.10[83] using the Slatkin's distance ($Rst$) matrix calculated in Arlequin. A Markov Chain Monte Carlo analysis of haplogroup structure was carried out using the program Structure 2.3.4[84]. YPredictor by Vadim Urasin v1.5.0 (http://predictor.ydna.ru/) was used for haplogroup prediction.

**Results and Discussion**

Our dataset has covered all the main haplogroups and almost all their sublineages in the Y chromosome phylogeny tree (Fig. 1a, Table 1S). Although the high mutation rates of STR markers make it difficult to construct phylogeny trees, the neighbor-joining tree of STR data (Fig.1b) shows a similar pattern as the trunk of Y chromosome haplogroup tree (Fig.1a). There are four main branches in the neighbor-joining tree, namely I, II, III, and IV. The three ancient haplogroup C, D, and E were clustered in branch I, representing the out of Africa migration. Haplogroup F, G, H, I, J were mainly clustered in branch II, which might indicate the possible population expansion in the Middle East and Europe. Haplogroup L, T, K, N, and O were clustered in branch III, representing the peopling of the Far East. Haplogroup P, Q, and R were mainly clustered in branch IV, demonstrating the population expansion from Central Eurasia and the migration through northern Eurasia into the Americas.

Obvious haplogroup divisions were also observed in the neighbor-joining tree, in which haplogroup R1b1+s (+s means certain haplogroup and its sublineages), haplogroup Q1+s, haplogroup G2a1+s, haplogroup J2+s, haplogroup E1b1b1+s, haplogroup D+s, and haplogroup E1b1a1+s were clustered tightly together, demonstrating the specific or even exclusive STR haplotypes of those haplogroups. However, haplogroups A, B were scattered in the tree, probably due to the high diversification among different sublineages of the two oldest haplogroups. Haplogroup C+s were mainly clustered with haplogroups D and E. Haplogroup C1 and C3 showed strong affinity with haplogroup E1b1b1, however, C2a* and C2a1 tended to cluster with E1b1a1. The three main subclades of haplogroup O, O1, O2, and O3, tended to be clustered together. Haplogroup O1a+s showed very strong affinity with haplogroup N* and N1c. Haplogroup O2a1 and O2a1a fell out the scale of STR patterns of haplogroup O. Haplogroup O2b* and O2b1a were clustered with T* and T1a*, and also showed affinity with O3a* and O3a1c*. Haplogroup O3a2+s formed a tight cluster, indicating high similarities between those lineages, although haplogroup L+s have also been placed in the O3a2 cluster. Haplogroup P was clustered with haplogroup Q+s in a separated small branch. Haplogroup R1a1 were clustered with E1b1a1 and C2a in branch I, while its sister clade R1b1+s were grouped with R2 and M1b in branch IV.

The neighbor-joining tree based on pairwise comparisons gives an overall clustering pattern of the worldwide haplogroups. However, the results at haplogroup level could be misleading because of the highly diversified STR haplotypes within haplogroups, especially in very ancient lineages. Here, we used Structure software to show the STR haplotype patterns among haplogroups at individual level (Fig 2). We also used YPredictor to infer haplogroup for each haplotype and then compared the inferred haplogroups with the genotyped haplogroups to estimate the error rates. The most ancient lineage A00 also has the exclusive STR haplotypes. The haplotypes of haplogroup A1a and A1b1b2b show similarities with haplogroup DE and E1b1a1, thus, about 30% of A1a and A1b1b2b samples were mistaken as DE in YPredictor. Haplogroup B+s are probably the most diverse clades, sharing similar haplotypes with various haplogroups, such as haplogroups I2a1, R1a1, D2a, E1b1b1, and L. Actually, only 18% of haplogroup B samples could be successfully inferred, and 26% were mistaken as I2 or IJ, 21% were assigned as haplogroup R in YPredictor. Similar to haplogroup B, the haplotypes of paragroup F*, H*, and K* are also too diverse to be used in haplogroup prediction. Most haplotypes of haplogroup C1 are similar to those of E1b1b1 and 22% of C1 samples were mistaken as E1b1b1 in prediction. Similarly, haplogroup C2 and its sublineages C2a and C2a1 shared most haplotypes with E1b1a1 and therefore 37% of those C2 samples were mistaken as E1b1a1. The haplotype pattern of haplogroup H1 and H1a is very similar to that of haplogroup J, resulting in erred assigning of about 20% of H1 and H1a samples as J in prediction. Haplogroup I*, I1*, and I1a1b1 share some haplotypes with haplogroup G+s. The haplotypes of haplogroup L+s are similar to those of haplogroup O3a2c1* and O3a2c1a. The haplotypes of O1a+s bear some similarity to those of haplogroup N. Similarly, haplotypes of haplogroup O3a1c show some similarity to O3a2b, and M1a and M1b are similar to O3a2+s. Those haplotype sharing or similarities among different haplogroups often mislead us in haplogroup prediction. On the contrary, haplogroup D2a, D3a, O2a1, O2b, R, and Q have haplogroup specific haplotypes and could be predicted by STR with high accuracy. It is still worthy to note that the affinity between haplogroup D and E might confuse the two in some cases.

The purpose of our paper is not to address the quality of the haplogroup prediction software, as no algorithms could be powerful enough to distinguish the same or very similar haplotypes and assign them into different haplogroups. The convergence of Y chromosome STR haplotypes among different haplogroups has compromised the accuracy of haplogroup prediction. For samples with ambiguous STR haplotypes, typing SNPs is the only reliable method to determine the haplogroups.

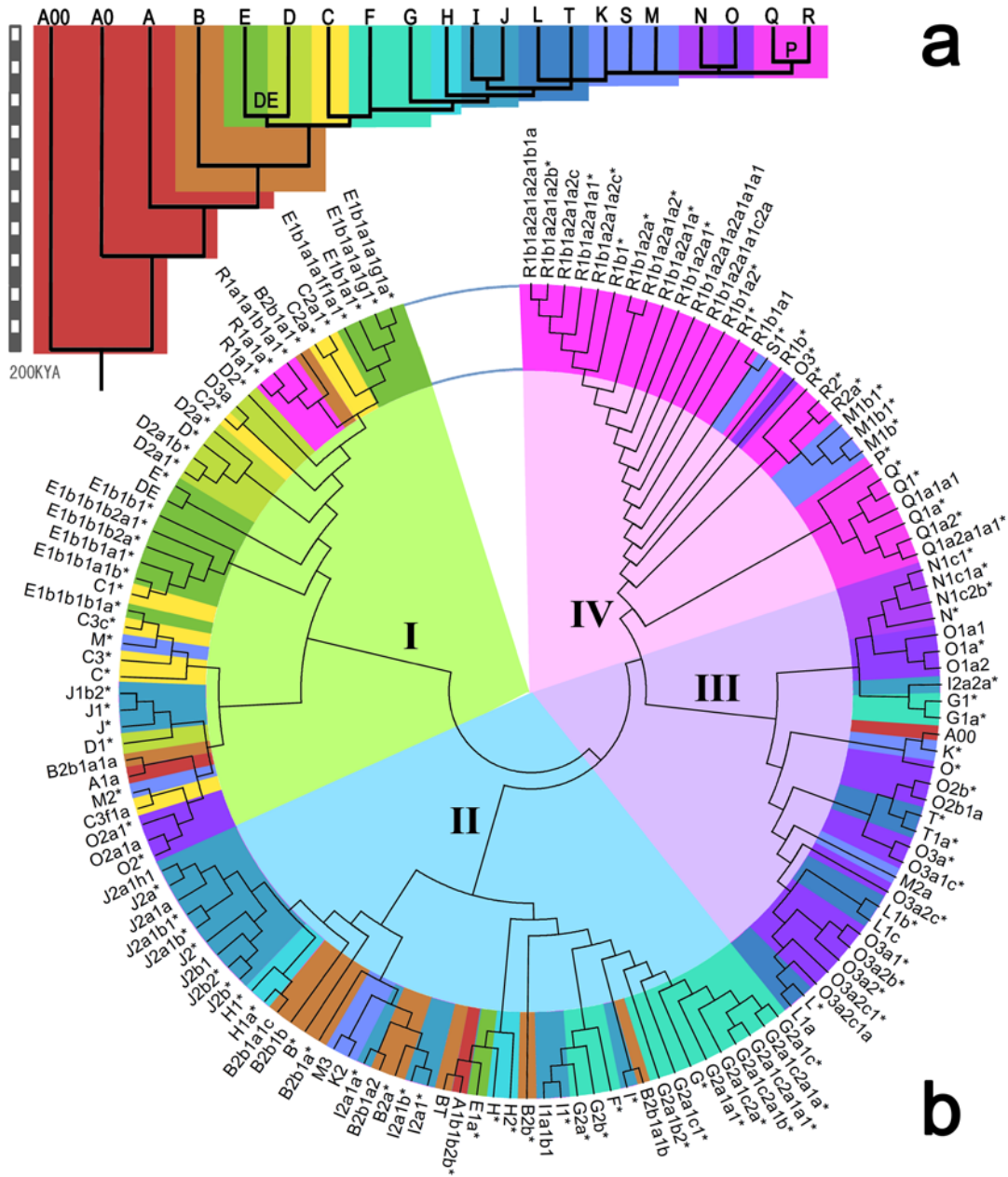

Fig.1 (a) The trunk of the Y chromosome haplogroup tree; (b) Neighbor-joining tree of 146 Y chromosome haplogroups based on Rst distance of 10 commonly used STRs (DYS19, DYS389I, DYS389II, DYS390, DYS391, DYS392, DYS393, DYS437, DYS438, and DYS439).

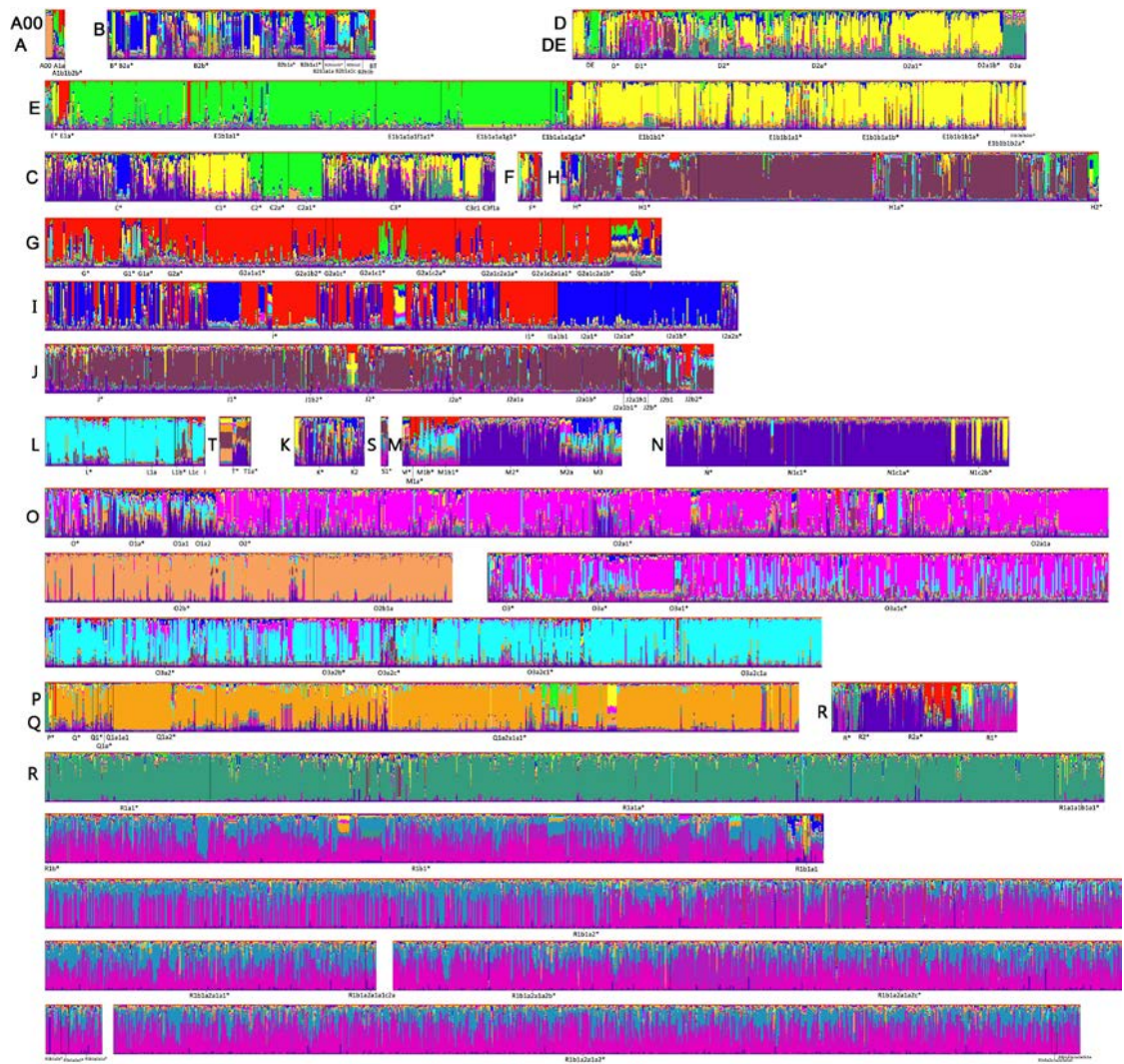

Fig.2 STR structure (K=14) of 146 Y chromosome haplogroups using 10 STRs as in Fig.1b

**Acknowledgments**

This work was supported by the National Excellent Youth Science Foundation of China (31222030), National Natural Science Foundation of China (31071098, 91131002), Shanghai Rising-Star Program (12QA1400300), Shanghai Commission of Education Research Innovation Key Project (11zz04), and Shanghai Professional Development Funding (2010001).

**Supplementary Material**

Table S1-Y SNP and STR dataset